# Ultrafast decoupling of quasiparticles and spin fluctuations in superconducting cuprates

Yuto Taniguchi[1], Ryo Kato[1], Tatsuya Amano[1], Hirotake Itoh[1!], Yohei Kawakami[1], Yuto Nakamura[2], Hideo Kishida[2], Christian Bernhard[3], Jure Demsar[4], Takahiko Sasaki[5], Terukazu Nishizaki[6], Kenji Yonemitsu[7], and Shinichiro Iwai[1*]

[1]*Department of Physics, Tohoku University, Sendai 980-8578, Japan*

[2]*Department of Applied Physics, Nagoya University, Nagoya 464-8603 Japan*

[3]*Department of Physics and Fribourg Center for Nanomaterials*

*, University of Fribourg, CH-1700 Fribourg, Switzerland*

[4]*Institute of Physics, Johannes Gutenberg University Mainz, 55128 Mainz, Germany*

[5]*Institute for Materials Research, Tohoku University, Sendai 980-8577, Japan*

[6]*Department of Innovative Mechanical and Electrical Engineering, Kyushu Sangyo University, Fukuoka, 813-8503, Japan*

[7]*Department of Physics, Chuo University, Tokyo 112-8551, Japan*

*Corresponding author Email: s-iwai@tohoku.ac.jp

! present address: *Department of Physics and Astronomy, Kwansei Gakuin University, Sanda 669-1330, Japan.*

**Abstract**

**Understanding how quasiparticles are generated following a rapid quench of superconductivity in high-$T$c cuprates is a key unresolved problem in nonequilibrium superconductivity. Here, we resolve these processes in optimally doped YBCO [$YBa_2Cu_3O_y$ (y~6.94, $T_c$=92K)] using broadband (0.16−4.1 eV, ~100 fs) and nearly single-cycle (6 fs) transient reflectivity spectroscopy. We show that within a few femtoseconds, enhanced electron–electron Umklapp scattering dominates, signaling a transient modulation of long-range Coulomb interactions on the eV scale. This regime is followed by a rapid suppression of the scattering rate of the mid-infrared absorption associated with carriers dressed by spin fluctuations. We attribute this observation to an ultrafast decoupling of quasiparticles from the spin-fluctuation background, occurring on a ~90 fs timescale set by the inverse optical gap. These findings reveal the correlated many-body dynamics underlying quasiparticle generation in cuprates and provide further clues for unconventional pairing mechanism.**

## I. Introduction

Thus far, research on the ultrafast dynamics of quasiparticles generated by the photoexcitation in high-$T_c$ superconducting (SC) cuprates has focused on the low-energy range [1-11]. This appears to be reasonable, considering the SC gap energy of several tens of meV and the fact that these studies focused on the dynamics of the gap (condensate). However, in strongly correlated systems, the emergence of quasiparticles (upon suppressing SC) is accompanied by spectral changes over a much larger energy range (~ eV) [1-3, 12-19], reflecting changes in the microscopic electronic structure [20-23]. In fact, the importance of adopting the three-band Hubbard model (Emery model [24]) from the perspective of charge-transfer (CT) energy being much smaller than the Hubbard $U$, covalency of bonds, or (normal-state) coherence has been recognized [25-28]. Moreover, theories suggest that long-range Coulomb interactions need to be considered when attempting to explain charge responses below and above $T_c$ across a wide energy scale [29, 30].

The permittivity spectrum of SC cuprates [Fig. 1, optimally doped $YBa_2Cu_3O_y$ (y~6.9, $T_c$=92K)] [31-35] is characterized by several contributions between the free carrier Drude response (below 0.1 eV) and the CT band (~ 3 eV) [see supplemental marerial (1) [36]]: i) The mid-infrared absorption (0.1~1 eV, hereafter the mid-IR band), reflecting carriers dressed by spin fluctuations [20, 37-44], and the electronic excitations near ii) 1.4 eV and iii) 1.9 eV. While the latter is attributed to the Zhang–Rice singlet (hereafter ZRS band) [20, 25], the origin of the 1.4 eV band is still debated [32]. Given the fact that the peak at ~1.4 eV matches the peak in the dielectric loss function [31], it has been associated with long-range Coulomb interactions and is found to be sensitive to the SC transition [29, 30].

Much discussion has been devoted to changes in the optical excitation spectra across $T_c$

of optimally doped, underdoped and overdoped cuprates. Either increases or decreases in the kinetic energy (accompanied with decreases or increases in the interaction energy) have been debated for the thermally driven transition between the SC and normal (pseudogap) states, raising debates on the mechanism of high-$T_c$ superconductivity in the cuprates [34, 45-48]. The nonequilibrium spectral weight transfer upon the photoinduced quasiparticle generation, has also been investigated by measuring transient reflectivity ($\Delta R/R$) spectra in $Bi_2Sr_2Ca_{0.92}Y_{0.08}Cu_2O_{8+\delta}$ (Y-Bi2212) [17] and in optimally doped $YBa_2Cu_3O_y$ [18]. However, due to limitations in the accessible spectral range (1–2.2 eV [17], 1.4–2.6 eV [18]) and temporal resolution (ca. 200 fs), the initial-stage dynamics of photoinduced quasiparticle generation remain unclear.

In this study, we investigated the generation dynamics of quasiparticles in the SC state of $YBa_2Cu_3O_{6.94}$ ($T_c$=92K) by a $\Delta R/R$ measurement over a broad spectral range (0.16–4.1 eV), as well as using time-resolved measurement employing ultrashort near to mid infrared (0.5–1 eV) 6 fs pulses. Comparison with experiments performed in the normal state [295 K, see Supplemental material (2) [36] for details] and the published equilibrium optical spectroscopy data [31-35] allows identifying the underlying processes. Our results show that photoexcitation results in an enhancement of the electron-electron (e-e) Umklapp scattering within a few femtoseconds (the Umklapp scattering was found [29, 30] to be suppressed in the SC state). This is followed by a decrease in the scattering rate of the mid-IR band occurring on the time scale of 90 fs, reflecting ultrafast decoupling of the quasiparticles from the spin-fluctuation background. Finally, the photoinduced Hartree shift of the CT energy has been identified by detecting the changes in the ZRS band.

## II. Methods

### A. Materials synthesis and characterization

$YBa_2Cu_3O_y$ single crystals were grown by the self-flux method using yttria crucibles [49-52]. The as-grown crystals were annealed at 450 °C for 7 days under an oxygen flow atmosphere, and naturally untwinned single crystals were carefully selected under a polarized microscope. The superconducting transition temperature $T_c$ was determined by the magnetization measurement using SQUID magnetometer (MPMS3, Quantum Design). The temperature dependence of the magnetization $M(T)$ shows sharp superconducting transition at $T_c = 92.0$ K. The oxygen content $y$ was estimated to be $y = 6.94$ according to the relation between a doping level and an annealing temperature [49, 50]. Untwinned single crystals prepared by the same method clearly exhibit a first-order vortex-lattice melting transition [51, 52], indicating that samples in this study are high-quality crystals with less disorder.

### B. Transient reflectivity measurements (100 fs)

For optical pump-probe spectroscopy for measuring transient reflectivity ($\Delta R/R$) spectra, the fundamental output from a Ti: Sapphire regenerative amplifier, operating at 1 kHz, 800 nm, with a pulse width of *ca*. 100 fs, was used for the excitation of optical parametric amplifiers (OPAs). The wavelength of an OPA for the pump pulse was set at 1400 nm (0.89 eV), whereas another OPA is used for the probe pulse covering the spectral range between 300 -8000 nm (0.16 -4.1 eV). Polarizations of the pump ($E_{pu}$) and probe ($E_{pr}$) pulses are parallel to the ***a***-axis of the single crystal of YBCO. The probe beam was focused on the center of the pumping area on the sample which is set in the conduction-type liquid-He cryostat (the window is a 3 mm-thick $BaF_2$ crystal. Polarization directions of the pump and probe beams are normal to each other. The diameters of pumping and probing areas were measured using a knife-edge and are 400-500 micron and 100-200 micron, respectively. Then the reflected beam from the sample was detected by Si (300-1000 nm), InGaAs (1000-2200 nm) or HgCdTe (2500-8000 nm) photodetector after passing through a monochromator. The pump-on and pump-off were alternately switched by the feed-back-controlled optical chopper, synchronized with the 1 kHz laser driver. Each probe shot was sampled using boxcar integrators. The observed intensity of respective shots was recorded in the PC to calculate $\Delta R/R$. Time resolution is 150-200 fs depending on the probing wavelength.

**C. Transient reflectivity measurements (6 fs)**

**6 fs infrared pulse generation.** The 6 fs pulse covering 1.2–2.3 μm, is generated by the method described in the references [53-56], i. e., a broadband infrared spectrum covering 1.2–2.3 μm is obtained by focusing a carrier-envelope phase (CEP) stabilized idler pulse (1.7 μm) from an optical parametric amplifier onto a hollow fibre set within a Kr-filled chamber (Femtolasers). Pulse compression is performed using both active mirror (19-ch linear micro machined deformable mirror) and the chirped mirror techniques. The pulse width is derived from the SHG autocorrelation. As described below, the sample is placed inside the cryostat. The pulse width is adjusted to be shortest at the sample position within the cryostat. That is, the pre-chirped pulse is injected into the cryostat to compensate for pulse broadening in the window.

**Pump-probe measurement.** We performed $\Delta R/R$ measurements for the single crystal of YBCO using a 6 fs pulse. After controlling the light intensity and polarization using a pair of wire-grid $CaF_2$ polarizers, the identical pump and probe pulses [polarizations of which are parallel to the ***a***-axis ($E_{pu}||a$, $E_{pr}||a$)] are focused onto the sample inside the cryostat (through the 3 mm $BaF_2$ window) using a 90-degree off-axis parabolic mirror. The spot diameters of the pump and probe lights are 170 μm. The probe light reflected from the sample is collimated by a lens, then split into two beams by a pellicle beam splitter and detected by separate detectors. One beam is detected directly by an InGaAs detector without passing through a spectrograph, while the other is detected by the same detector after passing through a monochromator. This detection system enables simultaneous measurement of the time evolution of $\Delta R/R$ (over the entire pulse spectrum) without wavelength resolution and the wavelength-resolved $\Delta R/R$ spectrum. Time resolution is 9 fs. Zero-time delay ($t_d$=0) is determined by using the first-order cross-correlation of the pump and probe lights with identical spectra, which modulates the $\Delta R/R$ signal in a configuration where the polarizations of the pump and probe lights are parallel. This first-order cross-correlation exhibits an interference pattern, allowing the maximum interference peak corresponding to the time origin in the pump-probe experiment to be identified with an accuracy better than 1 fs (for the wavelength of 1.6 micron with the period of oscillating electric field of 6 fs) under ideal conditions, but typically ca. 1-3 fs. However, this accuracy also depends on the visibility of the interference pattern modulating the pump-probe signal; the visibility strongly depends on the surface roughness. The conversion from the detection signal to $\Delta R/R$ is the same as the method used in the measurement system with the 100 fs pulse.

## III. Results

### A. Transient reflectivity ($\Delta R/R$) measurement using 100 fs pulses over 0.16 -4.1 eV spectral range

Figure 2(a) shows the time evolution of the $\Delta R/R$ spectrum recorded at 10 K upon excitation of the mid-IR band (pump photon energy of 0.89 eV), measured using 100 fs probe pulses. The experiment was performed with excitation density ($I_{ex}$) of 0.04 mJ/cm$^2$, which is below the threshold for depletion of superconductivity [Supplemental Material (3) [36]]. The $\Delta R/R$ data are plotted as a function of pump-probe delay ($t_d$) and probe photon energy. The $\Delta R/R$ spectrum at $t_d$=0 ps is characterized by a positive peak centered around 1.2 eV, matching the plasma edge of the steady-state reflectivity [Fig. 2(b)], with negative tails on either side. As shown in Figs. 2(c), (d), the spectral range corresponding to the mid-IR band ($\hbar\omega$ < 1 eV) exhibits the ultrafast change of $\Delta R/R$ from negative [(c) $t_d$= 0 ps] to positive [(d) $t_d$=0.2 ps]. Furthermore, a broad $\Delta R/R$<0 region is detected above 1.5 eV with the dip near 1.9 eV and shoulder near 2.9 eV corresponding to the peaks of the ZRS band and the CT band, respectively. Figure 2(e) shows the time evolutions of $\Delta R/R$ measured at (I) 0.62 eV (mid-IR band), (II) 1.17 eV (peak of $\Delta R/R$), (III) 1.90 eV (ZRS band) and (IV) 2.92 eV (CT band). The ultrafast sign (< 200 fs) change in $\Delta R/R$ of the mid-IR band [from $\Delta R/R$<0 to $\Delta R/R$>0, see in Fig. 2(e)] reflects the generation process of quasiparticles. This early process has not been discussed in detail until now, while the relaxation process on the time scale of ca. 2 ps (seen across the entire spectral range) is similar to earlier studies in this compound [5, 6, 8, 12, 14-16, 18] and reflects the recombination of photogenerated quasiparticles back into the condensate. The observed unusual spectral dependence of the early time dynamics (< 200 fs) is in strong contrast with the investigations in the normal state [57, 58]. In the normal state, the time-evolution of spectral changes could be well accounted for by a rapid electron thermalization on the < 20 fs time scale has been reported followed by equilibration with bosonic degrees of freedom [57, 58]. Indeed, our broadband Δ*R/R* data at 295 K support eralier studies as shown in Supplemental material (2) [36]. It is noteworthy that the maximum value of Δ*R/R* in the SC state (~$3x10^{-2}$ at 1.2 eV, $t_d$=0.2 ps) is approximately one order of magnitude larger than that in the normal state (~$3x10^{-3}$ at 1.2 eV, $t_d$=0.1 ps, Fig. S1), consistent with earlier single-color studies [12-16].

B. **Δ*R/R* measurement using 6 fs pulses**

Figure 3(a) illustrates the time evolution of Δ*R/R* for the measurement utilizing 6 fs near to mid-IR pulses (used for both pump and probe) covering a large part of the mid-IR band (0.5–1 eV). A negative Δ*R/R* appears instantaneously within several femtoseconds considering the instrumental response time of 9 fs. This is followed by the change of Δ*R/R* from negative to positive at $t_d$~100 fs. The slight oscillatory modulation of Δ*R/R* is also observed on a longer time scale [Supplemental material (4) [36]], attributed to coherent phonons at 3.5 and 4.5 THz (the periods of which are 285 and 220 fs) [59-61].

A two-dimensional plot of the time evolution of the Δ*R/R* spectra [Fig. 3(b)] clearly shows the spectral change from negative (blue) to positive (red) during the first 200 fs. In particular, the spectral change in the first 10 fs shows that the reflectivity decrease (Δ*R/R*<0) is most pronounced around 0.6 eV [$t_d$=-10 fs, Fig. 3(c), gray circles], which is attributed to the bleaching of the mid-IR band. Subsequent changes in the spectrum are characterized by the growth of the negative peak centered around 0.8 eV until $t_d$=10 fs [violet circles in Fig. 3(c) and blue circles in Fig. 3(d)]. The spectral shape at $t_d$=10 fs is similar to the normalized Δ*R/R* spectrum ($t_d$=0 ps, measured using a 100 fs pulse) [blue crosses in Fig. 3(d)]. Thus, the build-up time of the 0.8 eV dip is finite and is estimated to be several (< 10) femtoseconds after the instantaneous bleaching [considering time resolution of 9 fs and the accuracy of zero time-delay ($t_d$=0) of 1-3 fs (see Methods **C**)]. Then, the $\Delta R/R$ spectrum evolves into the positive flat shape until $t_d$=200 fs, as shown in Figs. 3(e) and 3(f) [alongside the normalized $\Delta R/R$ spectra measured using 100 fs pulses, shown by green ($t_d$=0.1 ps) in 3(e), and orange ($t_d$=0.2 ps) crosses in 3(f)].

## IV. Discussion

### A. Analysis of ΔR/R spectra using Drude-Lorentz model

$\Delta R/R$ spectra are analyzed using the Drude-Lorentz model (DLM) – see Supplemental material (5) [36]. We employed a phenomenological DLM that does not assume physical interpretations, rather than the commonly used extended Drude model [62]. Despite the large number of parameters, the characteristic shape of the $\Delta R/R$ spectrum, with peaks or shoulders at the positions of characteristic spectral features observed already in equilibrium, and the Kramers-Kronig-constrained nature of data analysis, enable us to determine changes of parameters for each oscillator. The results of the analysis for $t_d$ = 0 ps and 0.2 ps are presented as the solid lines in Figs. 4(a) and 4(b), respectively. In this analysis, we reproduced the $\Delta R/R$ spectra by considering the changes in the Drude response, the mid-IR band, the 1.4 eV band, the ZRS band, and the CT band upon excitation [Supplemental material (5) [36]]. Excellent agreement with the measured spectra was obtained by allowing variations of oscillator strengths ($f$) and scattering rates ($\gamma$) only for the Drude response ($f_D$, $\gamma_D$), Mid-IR band ($f_{MIR}$, $\gamma_{MIR}$), 1.4 eV band ($f_{1.4\ eV}$, $\gamma_{1.4\ eV}$), ZRS band ($f_{ZRS}$, $\gamma_{ZRS}$) and CT band ($f_{CT}$, $\gamma_{CT}$). Just in the case of the 1.4 eV band we also allowed variation of the central frequency ($\Delta\omega_{1.4\ eV}$). To account for mismatches in the penetration depths of the pump and probe lights, a multilayer film model is employed [Supplemental material (6) [36]]. Figures 4(c)-4(i) show the extracted time evolutions of $\Delta\gamma_{MIR}$ and $\Delta f_{MIR}$ [(c) and (f)], $\Delta\gamma_{1.4\ eV}$ and $\Delta f_{1.4\ eV}$ [(d) and (g)], and $\Delta\gamma_{ZRS}$ and $\Delta f_{ZRS}$ [(e) and (i)]. A peak shift of the 1.4 eV band ($\Delta\omega_{1.4\ eV}$) is also shown in Fig.4(h).

Focusing on the 0.5–1.5 eV spectral range of Fig. 4(a), the noticeable peak at 1.2 eV and the negative tail below 1 eV at $t_d$=0 ps are mainly a result of the increase in $f_{1.4\ eV}$ [Fig. 4(g)] [Supplemental material (5)]. References [29, 30] reported that the effect of the Umklapp processes (associated with the long-range Coulomb interactions) on the charge response is suppressed below $T_c$. Taking this into account, the sharp peak in $\Delta R/R$ at 1.2 eV and the negative tail in 0.5-1 eV at $t_d$=0 ps [in Fig. 4(a)] are both attributed to photoinduced recovery of $f_{1.4\ eV}$, weakened upon cooling from the normal to the SC state due to the suppression of the e-e Umklapp-scattering [Supplemental material (7) [36]]. Indeed, such an increase in $f_{1.4eV}$ is absent in the normal state [Fig. S10(e) in Supplemental material (8) [36]], supporting this interpretation.

Another significant feature is that $\Delta R/R$ in the mid-IR (0.5- 1 eV) spectral range, which is negative at $t_d$=0 ps, changes sign within 0.2 ps. The positive $\Delta R/R$ at $t_d$=0.2 ps in 0.5-1 eV [Fig. 4(b)] is attributed to a decrease in $\gamma$ of the mid-IR band, $\Delta\gamma_{MIR}$ <0 [Fig. 4(c)] [Supplemental material (5) [36]]. It is noteworthy that observed $\Delta\gamma_{MIR}$ <0 contrasts with the usual case, where the scattering rate $\gamma$ normally increases as a result of the optical excitation and/or heating of the electronic system. Since the mid-IR band originates from the coupling of carriers with the fluctuating spin background, the observed spectral changes suggest ultrafast decoupling of the quasiparticles and their background spin fluctuations. This observation is consistent with the photoinduced generation of a pseudogap state (detected after >300 fs) with reduced spin fluctuations, as suggested by recent results obtained by transient angle-resolved photoemission spectroscopy [63]. The reduced $\gamma_{MIR}$ in the pseudogap state is further supported by the behavior observed in the steady-state optical conductivity of underdoped Bi2223 near the pseudogap temperature [64].

### B. Characteristic timescales related to ultrafast recovery of Umklapp scatterings and decoupling of quasiparticles and their background spin fluctuations

The ultrafast (< 10 fs) decrease in reflectivity observed in 0.55–0.9 eV is, as pointed out above, indicative of an increase in $f_{1.4\,\mathrm{eV}}$ [Fig. 4(g)] [Supplemental material (5) [36]]. We attribute this response to the photoinduced enhancement of the e-e Umklapp scattering, which is suppressed in the SC state [Supplemental material (7) [36]]. Here, the 6 fs experiment demonstrates that the corresponding spectral change in reflectivity occurs within a few femtoseconds. Considering that the enhancement of reflectivity near the plasma reflectivity edge (~1.2 eV) due to the recovered Umklapp scatterings originates from the long-range Coulomb interactions [29, 30], its energy scale is ~1 eV. Thus, its response time of a few femtoseconds is reasonable.

Furthermore, the subsequent spectral change in $\Delta R/R$ for 0.5-1 eV [Figs 4(a) and 4(b)] indicates the decrease in $\gamma_{MIR}$ [Fig. 4(c)] [Supplemental material (5) [36]]. The time required for this process is estimated at approximately 90 fs, derived from Figs. 3(b) and 3(e) as the time taken for $\Delta R/R$ to cross zero from negative to positive. This timescale characterizes the decoupling of the quasiparticles and their background spin fluctuations is characterized by 90 fs. Notably, the inverse of this timescale is approximately equal to the optical conductivity gap ($\hbar$/90 fs ~ 46 meV) [65].

## C. Photoinduced change in oscillator strength of ZRS band and Hartree shift of CT band

As shown in Fig. 4(i), $f_{\mathrm{ZRS}}$ decreases at 10 K (in the SC state), while it increases in the normal state [at 295 K, Fig.S10(g) in Supplemental material (8) [36]]. The opposite tendency between the behavior in the SC and the normal states can be readily inferred from the sign of Δ*R/R* in the corresponding spectral range near 1.9 eV [Figs. 2(c),(d) or Figs. S1(e),(f) at 10 K and Figs. S1(b),(c) at 295 K]. We attribute this to the photoinduced Hartree shift.

The Hartree shift arises from changes in the hole-density distribution at the Cu 3*d* and O 2*p* orbitals as schematically indicated in Figs. 5(a), (b), and effectively modulates the CT transition energy between these orbitals through the Coulomb repulsion. When superconductivity is suppressed due to photoexcitation, holes are expected to be transferred from the *p* to *d* orbitals [Supplemental material (9) [36]] [66-69]. On the other hand, in the normal state, photoexcitation results in charge reshuffling [70], i.e., holes are expected to be transferred from the *d* to *p* orbitals. Thus, the photoinduced Hartree shift should result in opposite changes of the CT energy in the SC and normal states, i.e., upon photoexcitation the effective CT energy $\Delta_{pd}^{\mathrm{eff}} \equiv \varepsilon_p - \varepsilon_d + U_{pd}(n_d - n_p)$ is increased (blue shifted) in the SC state [Compare Fig. 5(b) with Fig.5(a)] and decreased (red shifted) in the normal state.

Because of the low excitation efficiency of the CT band by the 0.89 eV pump, the amplitude of the energy shift should be of the order of 1 meV or less [Supplemental material (10) [36]]. On the other hand, theoretical calculations based on exact diagonalization [25] did reveal a correspondence between the CT energy and $f_{\mathrm{ZRS}}$, accounting for the Hartree shift, as shown in Fig. 5(c). Specifically, an increase in $f_{\mathrm{ZRS}}$ [Fig. S10(g)] in the normal state corresponds to a decrease in the CT energy (violet filled circles in Fig. S10(i)). Thus, the observed decrease [SC state, Fig. 4(i)] and increase [normal state, Fig. S10(g)] in $f_{ZRS}$ in the experiments correspond to the Hartree-shift-induced changes in the CT energy.

## V. Summary and conclusions

In summary, the early-stage dynamics of the quasiparticle generation in optimally doped high-$T_c$ SC cuprates are in the SC state characterized by an increase in electron-electron Umklapp scatterings (the process that is reduced in the SC state) which occurs within a few femtoseconds. The subsequent decoupling of the quasiparticles and their background spin fluctuations occurs on the time scale of 90 fs [$\hbar$/90 fs ~ 50 meV (optical gap energy)]. Furthermore, the decrease in the spectral weight of the ZRS band is attributed to the Hartree shift of the CT energy, based on the theoretically derived relation between the ZRS band and the CT band. The mechanism by which the MIR peak narrows upon photoexcitation is significant and awaits theoretical analysis.

## Acknowledgements

This work was supported by JST CREST (JPMJCR1901) (S. I.), MEXT Q-LEAP(JPMXS0118067426)(S. I.), JSPS KAKENHI [JP20K03800(Y. K.), JP23K25797 (Y. K.), JP22H01149 (H. I.), JP25K17323 (T. A.), JP24K08236 (T. N.), JP23K25811 (T. S.)], DFG (TRR 288 – 422213477, B08) (J. D.) and SNSF 200021-214905 (C. B.).

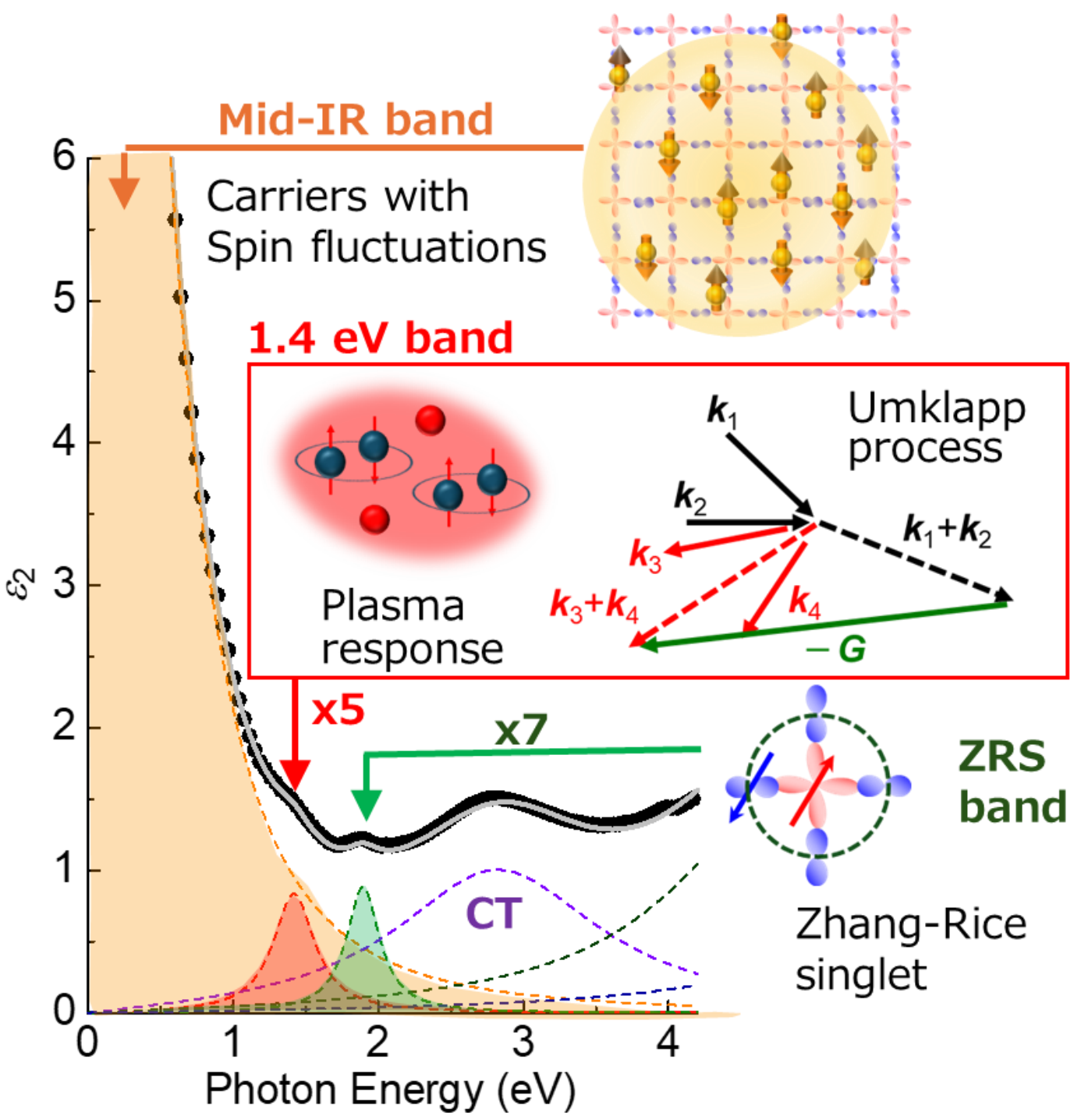

**Fig.1 Steady-state permittivity spectrum in optimally doped $YBa_2Cu_3O_{6.94}$ (20 K).** Steady-state permittivity spectrum in optimally doped $YBa_2Cu_3O_{6.94}$ (20 K). The origins of the mid-IR band (orange shade), 1.4 eV band (red shade x5), and ZRS band (green shade x7) are also schematically shown, i.e., carriers accompanied by spin fluctuations (mid-IR band), plasma responses associated with long-range Coulomb interactions (1.4 eV band) and Umklapp scatterings, and Zhang–Rice singlets (ZRS band).

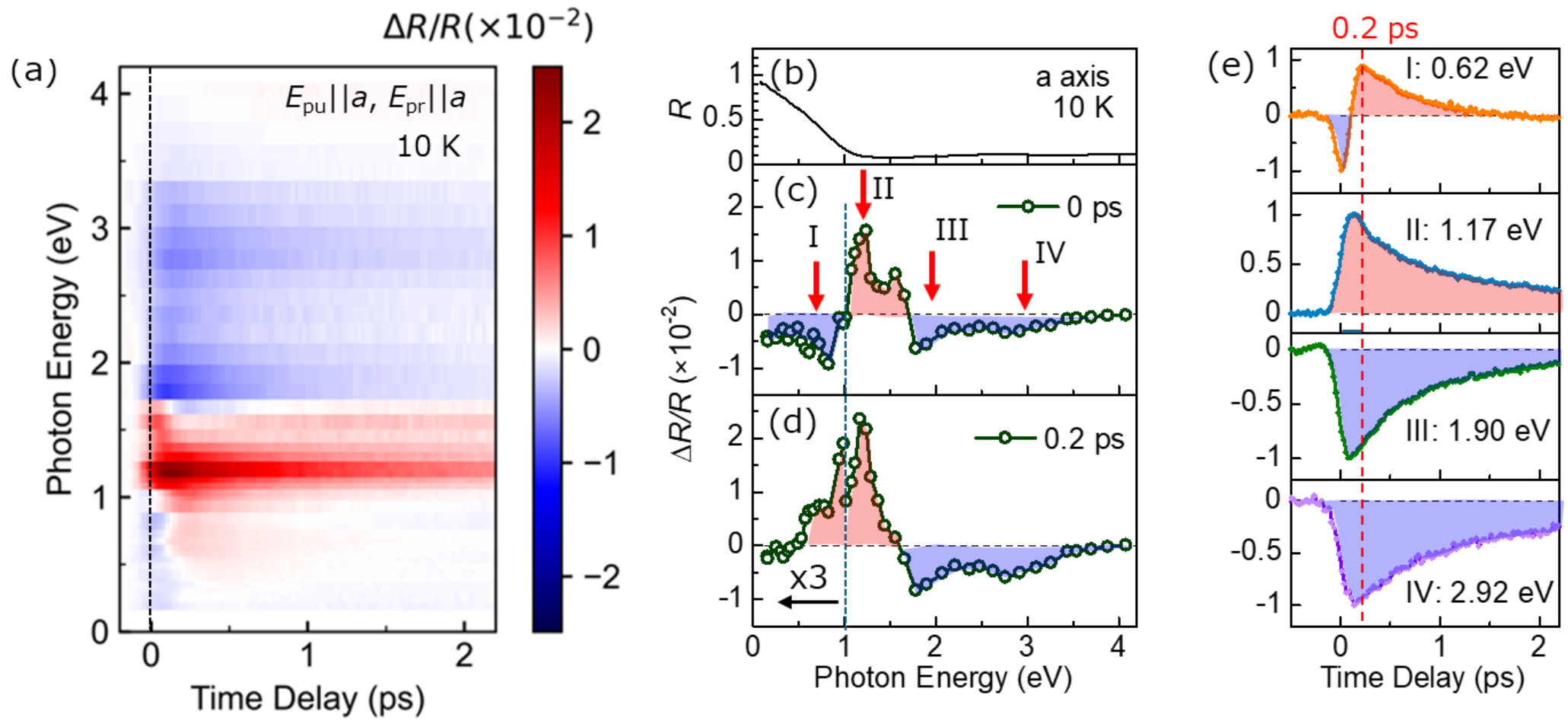

**Fig.2 Time evolution of transient reflectivity($\Delta R/R$) at 10 K.** (a) Two-dimensional plot of the time evolution of $\Delta R/R$ spectrum, measured using a 100 fs pulse (10 K, pump energy of 0.89 eV, $I_{ex}$= 0.04 mJ/cm²). Polarizations of pump ($E_{pu}$) and probe ($E_{pr}$) pulses are parallel to the ***a***-axis. (b) Steady-state $R$ spectrum and (c)(d) $\Delta R/R$ spectra measured at $t_d$=0 ps (c), 0.2 ps (d). The spectral range below 1 eV is shown multiplied by 3. (e) Time evolutions of $\Delta R/R$ measured at (I) 0.62 eV (mid-IR band), (II) 1.17 eV (peak of $\Delta R/R$), (III) 1.90 eV (ZRS band), and (IV) 2.92 eV (CT band).

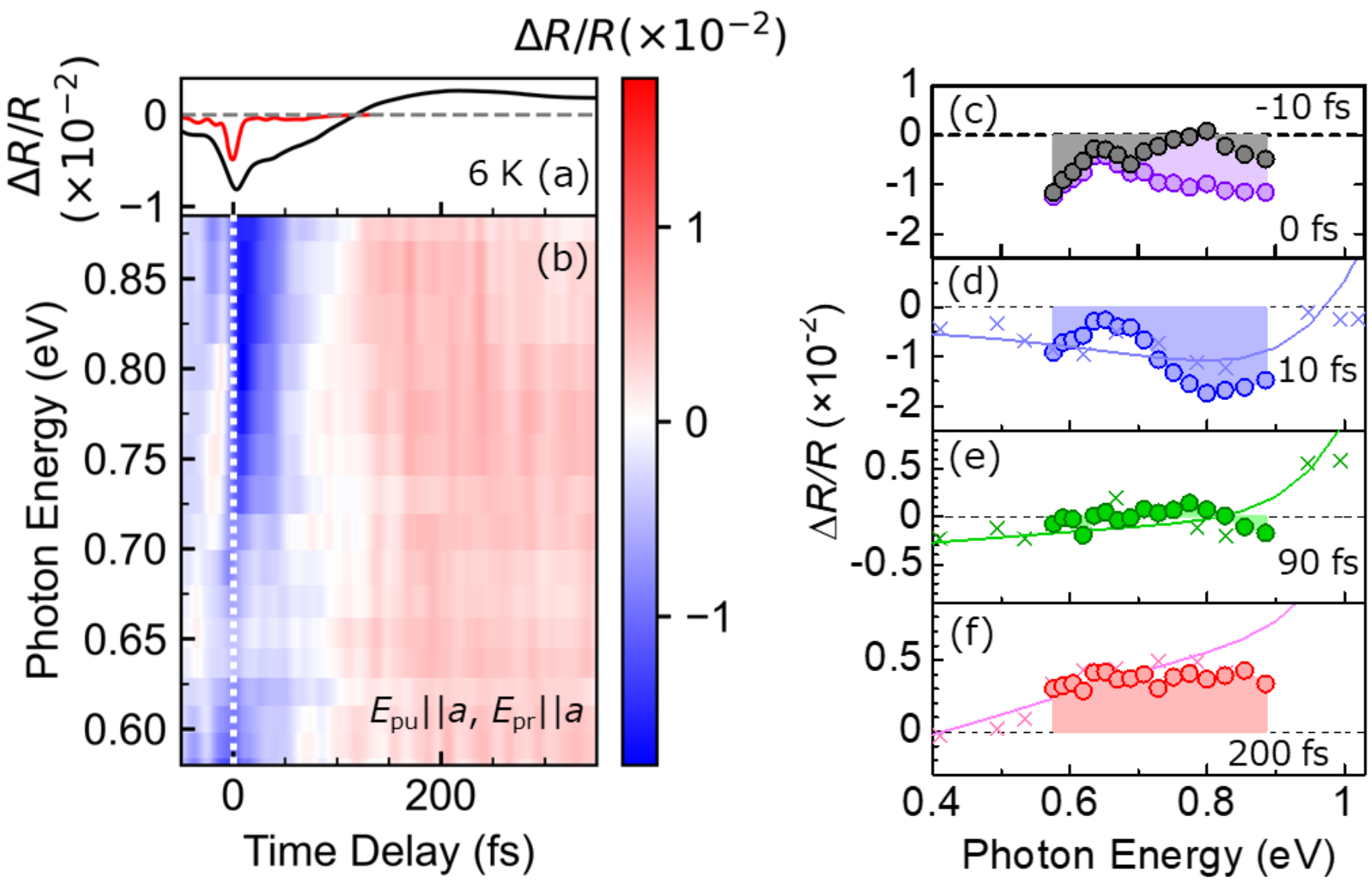

**Fig.3 Δ*R/R* spectra measured by using 6 fs pulse.** (a) Spectrally integrated time evolution of Δ*R/R* (The red curve shows the cross-correlation of pump and probe pulses) and (b) time evolution of Δ*R/R* spectrum, measured using a 6 fs pulse. Polarizations of pump ($E_{pu}$) and probe ($E_{pr}$) pulses are parallel to the ***a***-axis (6 K, $I_{ex}$= 0.1 mJ/cm²). (c)-(f) Δ*R/R* spectra at $t_d$=-10, 0 fs (c), 10 fs (d), 90 fs (e), 200 fs(f) are shown. The crosses and solid lines show the normalized Δ*R/R* spectra measured by 100 fs pulses and the results of the DLM analysis (shown in Figs. 2(c)(d) and 4(a)(b)), respectively.

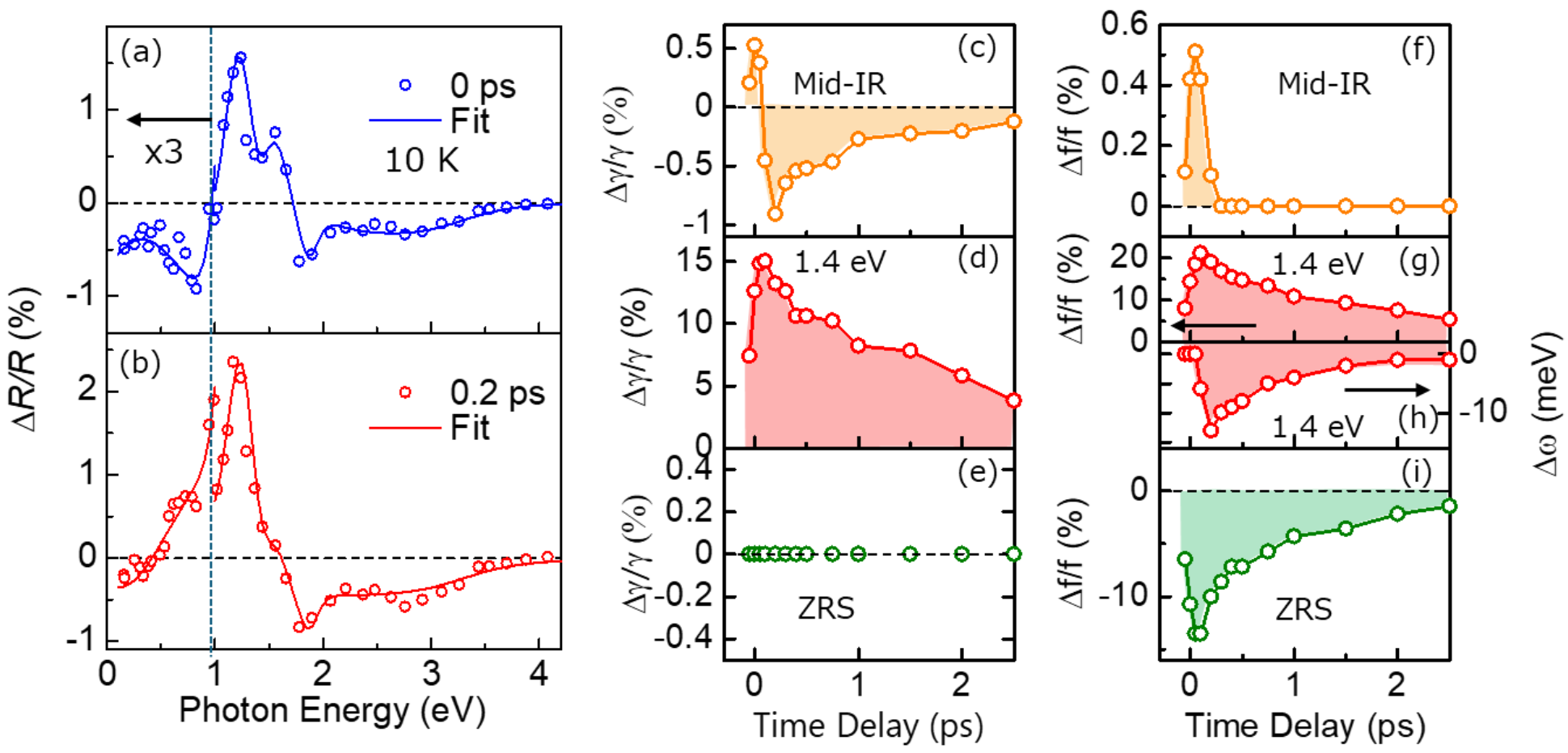

**Fig.4 Fig. 4 Results of Drude-Lorentz (DLM) analysis (10 K).**

(a)(b) Results of the DLM analysis for the $\Delta R/R$ spectra at $t_d$ = 0 ps (a) and 0.2 ps (b) are presented as the solid lines. (c)-(i) Time evolutions of $\Delta\gamma_{MIR}$ and $\Delta f_{MIR}$ [(c) and (f)], $\Delta\gamma_{1.4\text{ eV}}$ and $\Delta f_{1.4\text{ eV}}$ [(d) and (g)], and $\Delta\gamma_{ZRS}$ and $\Delta f_{ZRS}$ [(e) and (i)]. (h) Time evolution of the peak shift of the 1.4 eV band ($\Delta\omega_{1.4\text{ eV}}$).

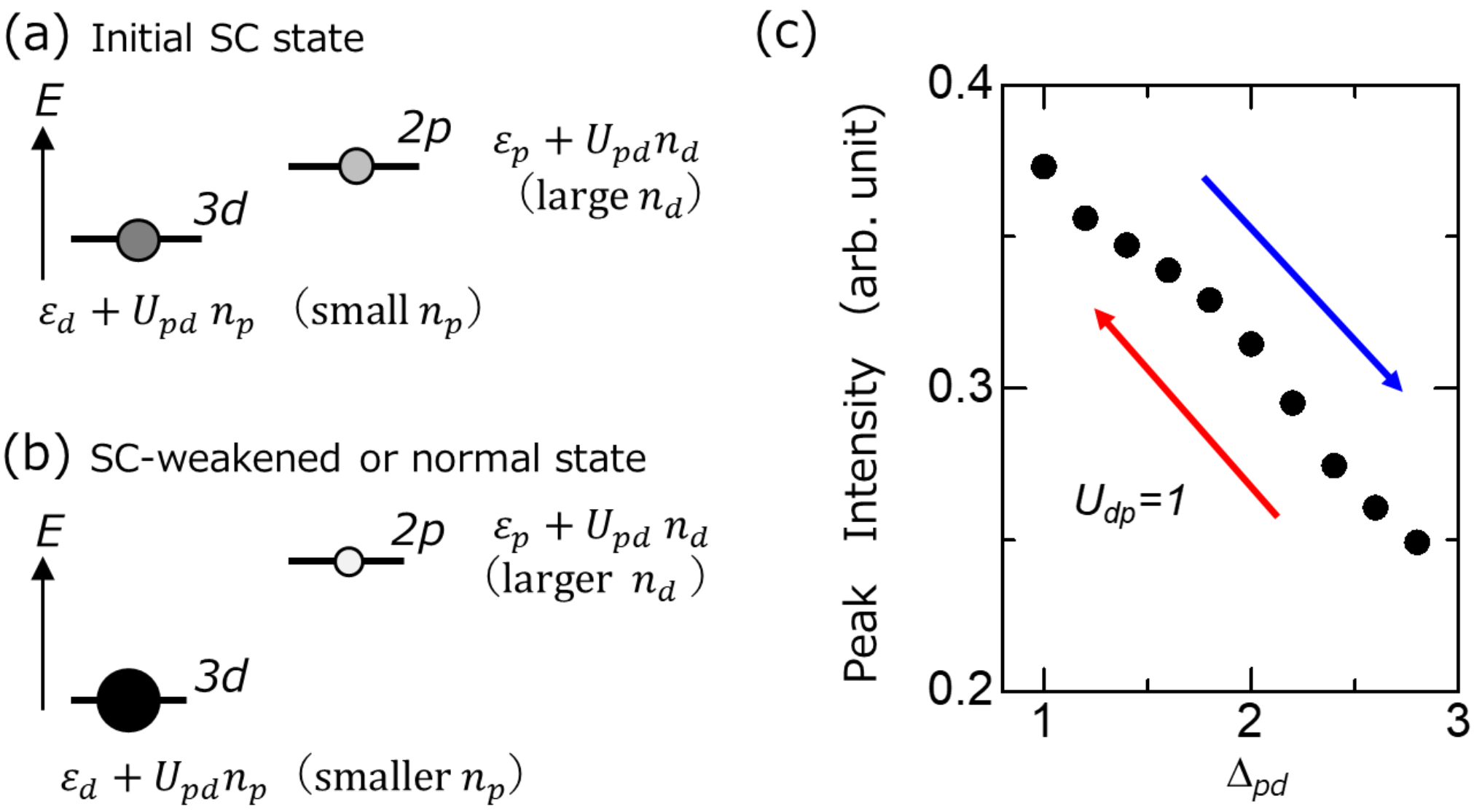

**Fig.5** (a)(b) Schematic illustrations of Hartree shifts of the CT energy in the SC state (a) and the normal state (b). The size of the circles indicate the hole densities of 3*d* and 2*p* orbitals. (c) Relation between the CT energy[$\Delta_{pd} = \varepsilon_p - \varepsilon_d$ with the on-site energy of the O 2*p* (Cu 3*d*) orbital $\varepsilon_p$ ($\varepsilon_d$) in the hole picture] and $f_{\mathrm{ZRS}}$ calculated based on the exact diagonalization [25]. The parameters and boundary conditions used here are the same as those in the 1-hole case of Fig. 12(b) in Reference 25 except for the CT energy.

## Supplemental Material:

# Ultrafast decoupling of quasiparticles and spin fluctuations in superconducting cuprates

Yuto Taniguchi[1], Ryo Kato[1], Tatsuya Amano[1], Hirotake Itoh[1!], Yohei Kawakami[1], Yuto Nakamura[2], Hideo Kishida[2], Christian Bernhard[3], Jure Demsar[4], Takahiko Sasaki[5], Terukazu Nishizaki[6], Kenji Yonemitsu[7], and Shinichiro Iwai[1*]

[1]*Department of Physics, Tohoku University, Sendai 980-8578, Japan*

[2]*Department of Applied Physics, Nagoya University, Nagoya 464-8603 Japan*

[3]*Department of Physics and Fribourg Center for Nanomaterials*

*, University of Fribourg, CH-1700 Fribourg, Switzerland*

[4]*Institute of Physics, Johannes Gutenberg University Mainz, 55128 Mainz, Germany*

[5]*Institute for Materials Research, Tohoku University, Sendai 980-8577, Japan*

[6]*Department of Innovative Mechanical and Electrical Engineering, Kyushu Sangyo University, Fukuoka, 813-8503, Japan*

[7]*Department of Physics, Chuo University, Tokyo 112-8551, Japan*

[*]Corresponding author Email: s-iwai@tohoku.ac.jp

[!] present address: Department of Physics and Astronomy, Kwansei Gakuin University, Sanda 669-1330, Japan

**Supplemental material (1)**

**Analysis of steady-state spectrum using Drude-Lorentz model (DLM)**

The steady-state permittivity ($\varepsilon$) spectrum [34] of optimally doped $YBa_2Cu_3O_y$(y=6.9) single crystal is analyzed using the Drude-Lorentz model (DLM) [71, 17]

$$\varepsilon(\omega) = \varepsilon_\infty - \frac{f_D \omega_p^2}{\omega^2 + i\omega\gamma_D} + \sum_{j=1}^{6} \frac{f_j \omega_p^2}{\omega_j^2 - \omega^2 - i\omega\gamma_j} \qquad \text{(eq.s1)}$$

, where $f_D + \sum_{j=1}^{6} f_j = 1$.

($f_D$, $f_j$, $\gamma_D$ and $\gamma_j$ are the oscillator strengths and scattering rates of the Drude response and the respective oscillators. $\omega_p$ is the plasma frequency). The steady-state spectrum comprises the Drude response, the mid-IR band ($j$=1), the 1.4 eV band ($j$=2), the ZRS band ($j$=3), the CT band ($j$=4) and interband transitions in the high-energy (>4 eV) range ($j$=5, 6). The results of the fitting analysis for 20 K are shown in Fig. 1 of the main text. The parameters used for the fitting are shown in Table S1. In the main text, $f_j$, $\gamma_j$ ($j$=1~4) are referred as $f_{MIR}$, $\gamma_{MIR}$ ($j$=1), $f_{1.4\ eV}$, $\gamma_{1.4\ eV}$ ($j$=2), $f_{ZRS}$, $\gamma_{ZRS}$ ($j$=3), $f_{CT}$, $\gamma_{CT}$ ($j$=4), respectively, for readability.

| | | 10 K | 295 K |
|---|---|---|---|
| Background permittivity | $\varepsilon_\infty$ | 2.16 | 2.16 |
| Plasma frequency | $\omega_p^2$ [eV$^2$] | 55.16 | 55.65 |
| Drude Response | $\gamma_D$ [eV] | 0 | 0.07 |
| | $f_D$ ($\times 10^{-2}$) | 3.11 | 2.59 |
| Mid-IR band | $\omega_1$ ($\omega_{MIR}$) [eV] | 0.1 | 0.1 |
| | $\gamma_1$ ($\gamma_{MIR}$) [eV] | 0.88 | 0.89 |
| | $f_1$ ($f_{MIR}$) ($\times 10^{-2}$) | 6.96 | 7.33 |
| 1.4 eV band | $\omega_2$ ($\omega_{1.4\ eV}$) [eV] | 1.4 | 1.43 |
| | $\gamma_2$ ($\gamma_{1.4\ eV}$) [eV] | 0.50 | 0.50 |
| | $f_2$ ($f_{1.4\ eV}$) ($\times 10^{-2}$) | 0.32 | 0.29 |
| ZRS band | $\omega_3$ ($\omega_{ZRS}$) [eV] | 1.89 | 1.86 |
| | $\gamma_3$ ($\gamma_{ZRS}$) [eV] | 0.27 | 0.3 |
| | $f_3$ ($f_{ZRS}$) ($\times 10^{-2}$) | 0.13 | 0.15 |
| CT Band O2p - Cu3d | $\omega_4$ ($\omega_{CT}$) [eV] | 2.94 | 2.93 |
| | $\gamma_4$ ($\gamma_{CT}$) [eV] | 1.75 | 1.85 |
| | $f_4$ ($f_{CT}$) ($\times 10^{-2}$) | 9.51 | 10.01 |
| 4.9 eV band O 2p - Ba 5d | $\omega_5$ [eV] | 4.87 | 4.87 |
| | $\gamma_5$ [eV] | 1.8 | 1.9 |
| | $f_5$ ($\times 10^{-2}$) | 23.11 | 23.29 |
| 6.5 eV band Cu3d, O2p - Ba, Y | $\omega_6$ [eV] | 6.5 | 6.5 |
| | $\gamma_6$ [eV] | 1.0 | 1.0 |
| | $f_6$ ($\times 10^{-2}$) | 56.86 | 56.35 |

**Table S1 Parameters used for DLM analysis of steady state permittivity spectra.** ($YBa_2Cu_3O_{6.9}$ single crystal, 20 K, 295 K [34]).

In this analysis, we used the steady state $\varepsilon_1$ and $\varepsilon_2$ data obtained from the ellipsometry measurement [34]. These $\varepsilon_1$ and $\varepsilon_2$ spectra are consistent with our reflectivity spectrum shown in Fig. 2(b) (in main text) and Fig. S1(d) in supplemental material (2).

Note: When analyzing the data in the superconducting state we consider the Ferrell-Glover-Tinkham sum rule (roughly obeyed up to 100 meV – [48]) and thus approximate the low frequency response with Drude.

**Supplemental material (2)**

**Δ*R/R* spectrum measurement in normal state (100 fs pulse)**

Figures S1(a) and S1(b), (c) show the steady-state reflectivity [Fig. S1(a)] and Δ*R/R* spectra [Figs. S1(b),(c)] measured at 295 K [$t_d$ = 0.1 ps (b) and 0.2 ps (c) ]. The spectral range above 2.5 eV is shown multiplied by 5. The Δ*R/R* spectrum shows a decrease in reflectivity below 1.1 eV (blue shaded) and an increase above 1.1 eV (red shaded). The spectral shape of Δ*R/R* measured at 295 K differs from that observed in the SC state [Figs.S1(e),(f)] in the sense that the sharp positive peak at ~1.2 eV is absent in the normal state. Furthermore, the characteristic spectral change in the SC phase, that is, the change from $\Delta R/R < 0$ to $\Delta R/R > 0$ occurring between $t_d = 0.1$ ps and 0.2 ps in the spectral region below 1 eV, is also absent in the normal state. The $\Delta R/R > 0$ in the spectral range between 1.5-2.5 eV in the normal state shows the opposite sign to $\Delta R/R < 0$ in the SC state. These opposite behaviors of the spectral structures near the ZRS band in the SC and normal states are related to the opposite Hartree shifts described in the discussion (Photoinduced change in oscillator strength of ZRS band and Hartree shift of CT band) of the main text [see also Supplemental materials (9), (10)].

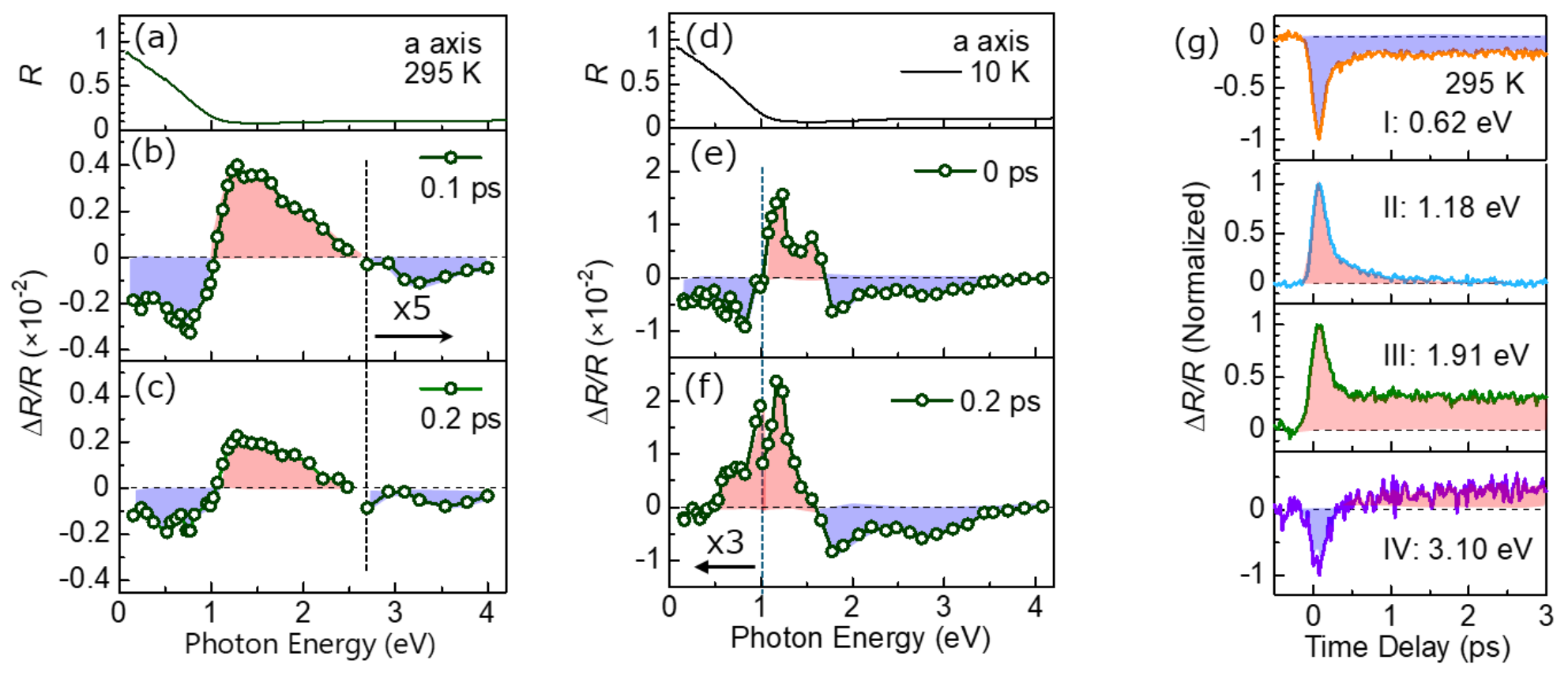

**Fig. S1 Δ*R/R* measured at 295 K and 10 K.** (a) Steady state *R* spectrum and (b)(c) Δ*R/R* spectra measured at 295 K. $t_d$=0.1 ps (b), 0.2 ps (c). (pump photon energy 0.89 eV, $I_{ex}$= 0.04 mJ/cm²). The spectral range above 2.5 eV is shown multiplied by 5. (d) Steady state *R* spectrum and (e)(f) Δ*R/R* spectra at 10 K [same as Figs. 2(b)~(d)]. Note that the signal amplitude in the SC state is approximately one order of magnitude larger [Δ*R/R*~3 x$10^{-2}$ at 1.2 eV ($t_d$=0.2 ps)] than that in the normal state [Δ*R/R*~0.3x$10^{-2}$ at 1.2 eV ($t_d$=0.1 ps)]. (g) Time evolutions of Δ*R/R* measured at 295 K for (I) 0.62 eV (mid-IR band), (II) 1.18 eV (peak of Δ*R/R*), (III) 1.91 eV (ZRS band), (IV) 3.10 eV(CT band).

Figure S1(g) shows the time evolution of $\Delta R/R$ measured at 295K for (I) 0.62 eV (mid-IR band), (II) 1.18 eV, (III) 1.91 eV (ZRS band), and (IV) 3.10 eV (CT band). At all probe photon energies, the time evolution of $\Delta R/R$ at 295 K shows an instantaneous build-up and decay with a time constant of ca. 150 fs, as well as long-lived (bolometric) response. The observed relaxation on the 150 fs timescale is consistent with earlier studies and is attributed to thermalization of carriers with bosonic excitations [57, 58]. These temporal profiles differ dramatically from those in the SC state, where recovery proceeds with the decay time constant of ca. 2 ps

Some parts of the spectrum (0.7–1.5 eV[57] and 0.6–1.9 eV[58]) have already been observed in the normal state. Those results have demonstrated that the thermalization of the electronic system occurs within 20 fs. The results presented here are essentially consistent with those earlier investigations, although the increase in the oscillator strength of the ZRS band [Supplemental material (5)] is observed in the present study.

**Supplemental material (3)**

**Excitation density dependence**

Excitation density ($I_{ex}$) of 0.04 mJ/cm² corresponds to the absorbed photon density of 9.8 x $10^{-4}$ /site. As seen in Fig. S2, $\Delta R/R$ shows linear response for $I_{ex}$< 0.06 mJ/cm$^2$.

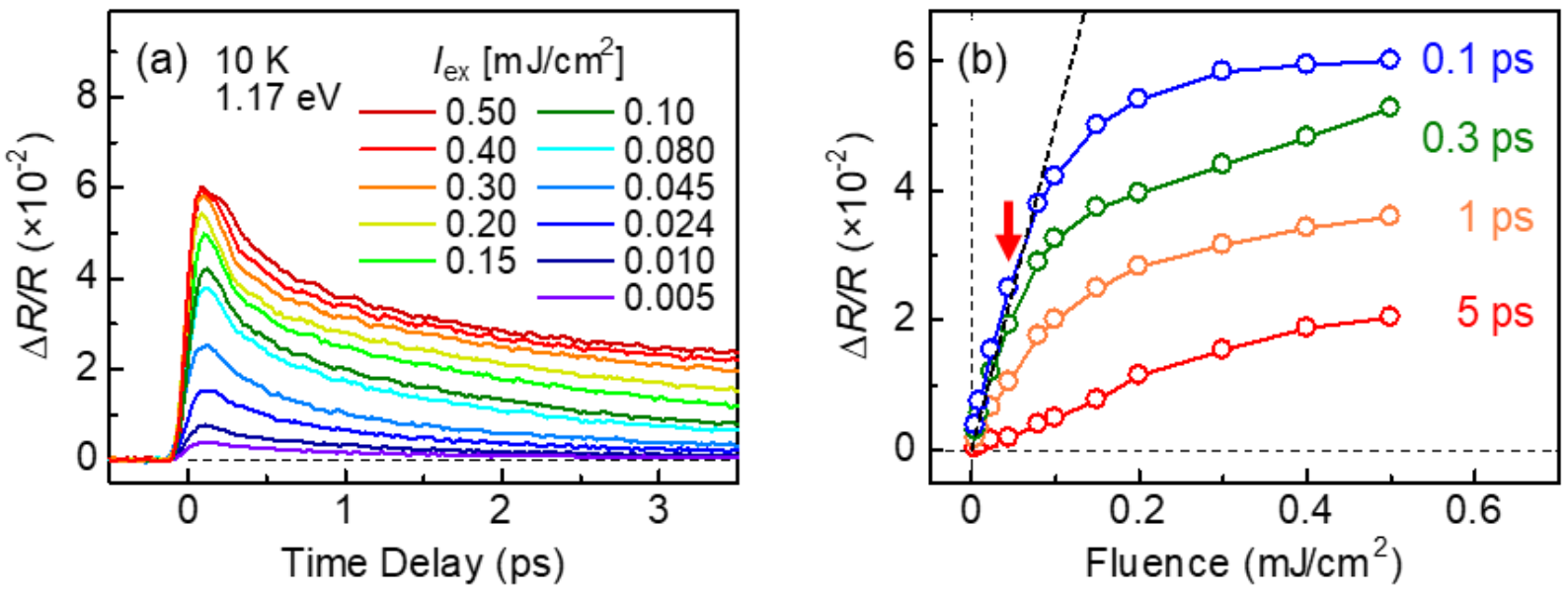

**Fig. S2 Excitation density dependence.** (a) Excitation density ($I_{ex}$) dependence of time evolution of $\Delta R/R$ measured at 1.17 eV (10 K). (b) $I_{ex}$ dependence of $\Delta R/R$ at $t_d$=0.1 ps. Linear response is confirmed below $I_{ex}$=0.06 mJ/cm$^{-1}$ (red arrow). Those at $t_d$=0.3, 1 and 5 ps are also shown.

**Supplemental material (4)**

**Oscillating structures on longer time scale**

Figure S3 shows the time evolution of $\Delta R/R$ measured using 6 fs pulses [same as in Fig. 3(a), but in a longer time window up to 700 fs]. The oscillatory modulation of $\Delta R/R$ is clearly seen on this time scale. These oscillatory components are attributed to the coherent phonons with the frequencies of 3.5 and 4.5 THz, which have been extensively discussed [59, 60, 18, 61] as described in the main text. This paper focuses on the generation process of quasiparticles in the <200 fs range, so we do not discuss these coherent phonons in further detail.

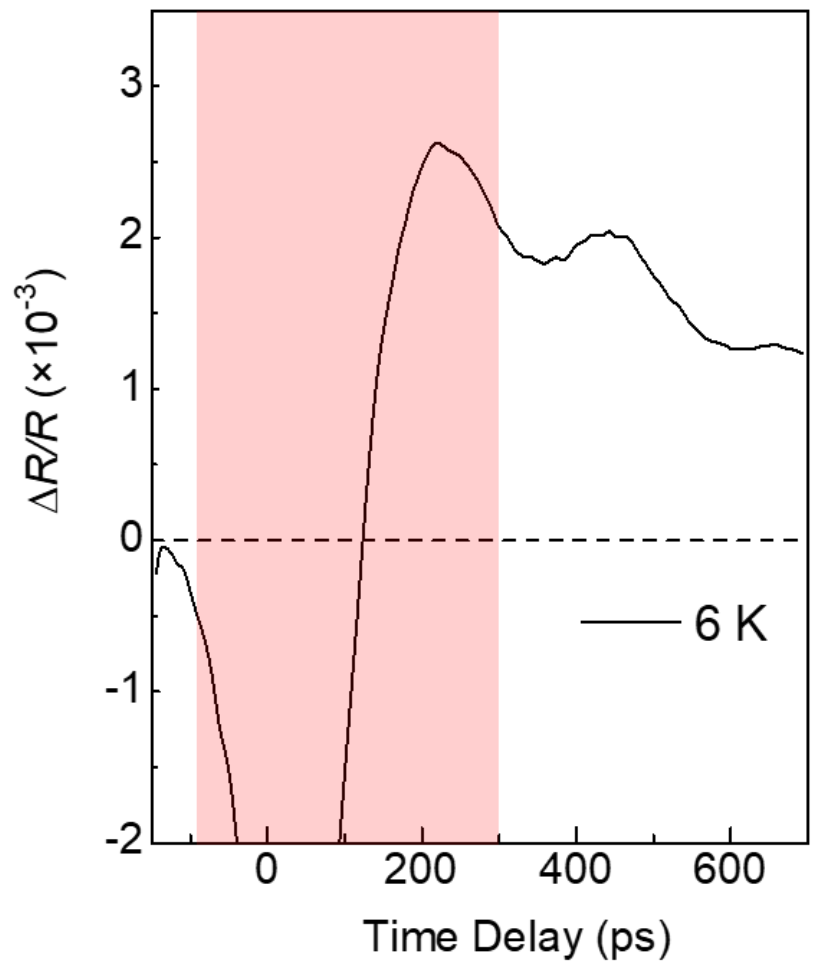

**Fig. S3 Time evolution of $\Delta R/R$ in longer time window.** The time domain shown in Fig. 3(a) of the main text is indicated by the red shaded area.

**Supplemental material (5)**

**Contribution from respective oscillators in DLM analysis**

**10 K**

The DLM analysis performed in this study is based on the Drude response and the six Lorentz oscillators as described by eq. s1 in the Supplemental material (1) (we assume the high frequency modes in the UV range are not affected by photoexcitation). Despite the large number of parameters, the characteristic shape of the $\Delta R/R$ spectrum with peak or shoulder structures A-F in the SC state and the Kramers-Kronig constrained nature enable us to determine the parameters for each oscillator. Figures S4(a),(d) show the results of the analysis at $t_d$ = 0 ps and 0.2 ps, respectively, for the $\Delta R/R$ spectra at 10 K, which are presented in Figs.4(a),(b), respectively.

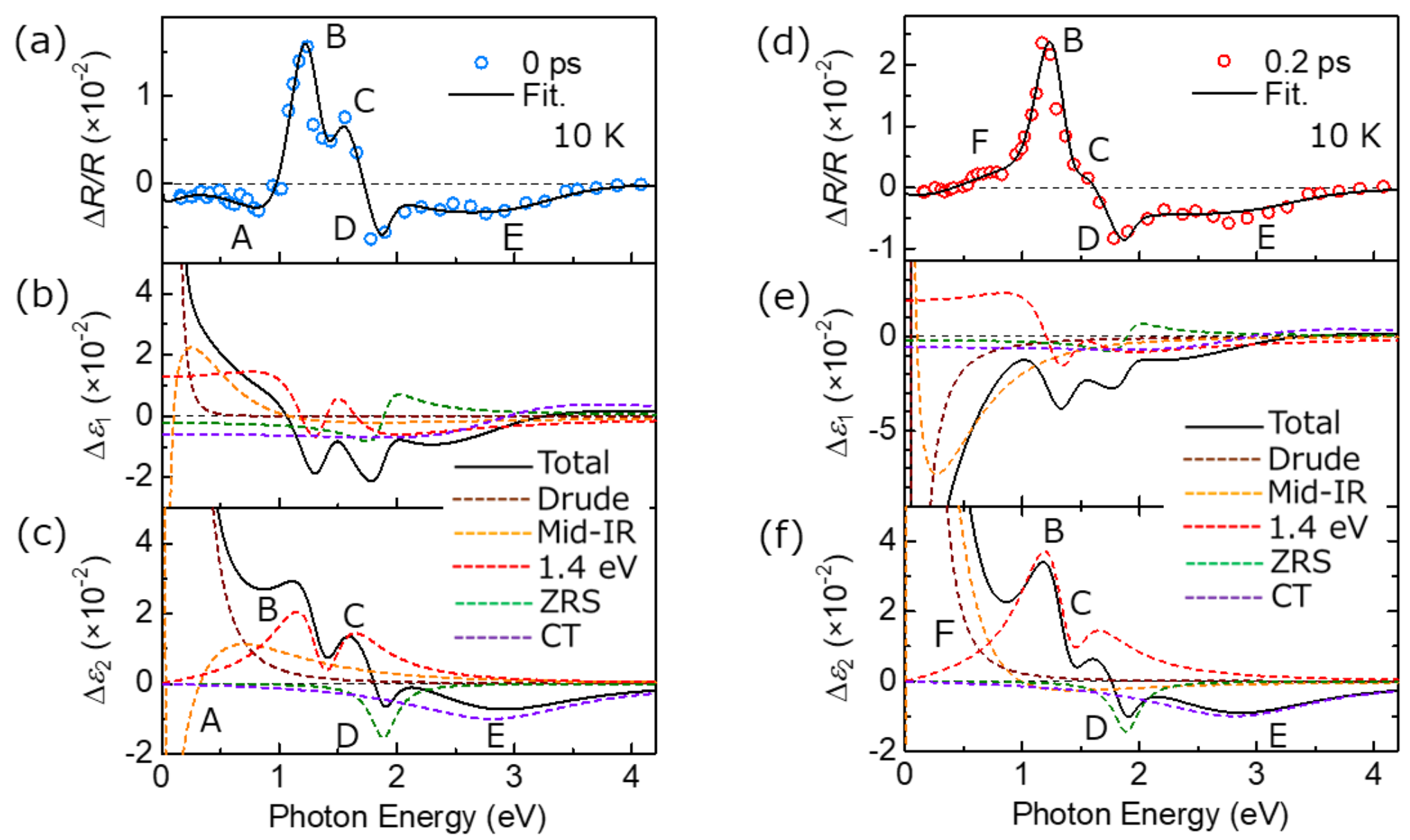

**Fig. S4 Detaila of spectral analysis (10 K).** (a)(d) Results of the analysis for the $\Delta R/R$ spectra at 10 K [$t_d$ = 0 ps (a) and 0.2 ps (d)].
For A-F, refer to the text below. (b)-(f) Contribution of each oscillator to the shape of $\Delta\varepsilon_1$ (b)(e) and that of $\Delta\varepsilon_2$ (c)(f), i.e., the respective spectra were calculated by varying $f$ and $\gamma$ of a single focused oscillator. See text.

The parameters used for the fitting are shown in Table S2. Figures S4(b),(c),(e),(f) show the contribution of each oscillator to the shape of the transient change in the permittivity $\Delta\varepsilon_1$ (b),(e) and that in $\Delta\varepsilon_2$ (c),(f), i.e., the respective spectra were calculated by varying $f$ and $\gamma$ of a single focused oscillator [either the Drude response ($f_D$, $\gamma_D$), the mid-IR band ($f_{MIR}$, $\gamma_{MIR}$), the 1.4 eV band ($f_{1.4\,eV}$, $\gamma_{1.4\,eV}$), the ZRS band ($f_{ZRS}$, $\gamma_{ZRS}$) or the CT band ($f_{CT}$, $\gamma_{CT}$)].

For Fig. S4(b)-(f), the contribution from each oscillator can be summarized as follows:

(i) The sharp peak structure (~1.2 eV) (B) in $\Delta R/R$ is due to the increased $f_{1.4\,eV}$ (red dashed curve), whereas the subpeak structure (~1.5 eV) (C) is due to the increase in $\gamma_{1.4\,eV}$.

(ii) The dip of reflectivity decrease (~1.9 eV) (D) is due to the decreased $f_{ZRS}$ (green dashed curve).

(iii) The broad decrease in reflectivity around 2–4 eV (E) is due to a decrease in $f_{CT}$ (violet dashed curve).

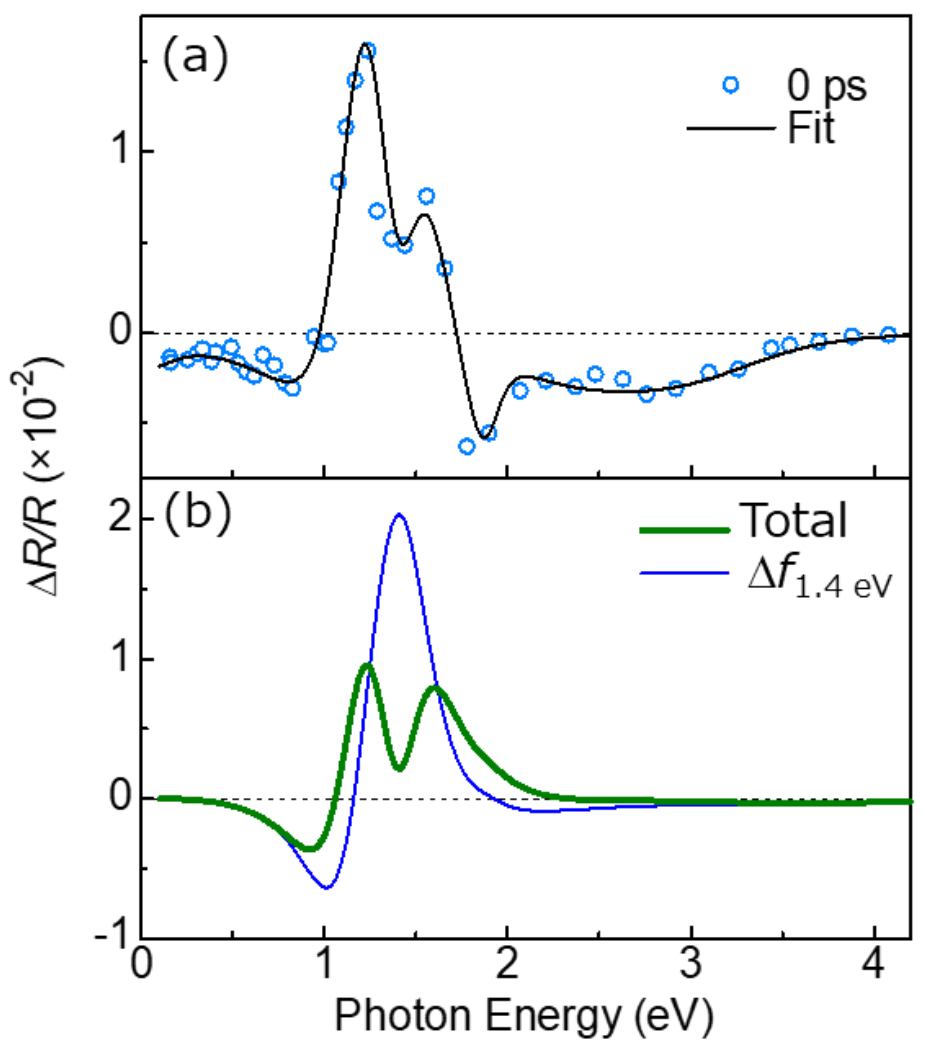

**Fig. S5 Contributions of $\Delta f_{1.4\,eV}$ and $\Delta\gamma_{1.4\,eV}$.** (a) Results of the analysis for the $\Delta R/R$ spectrum measured at 10 K [$t_d = 0$ ps same as Fig. S4(a)]. (b) Contributions of $\Delta f_{1.4\,eV}>0$ (blue curve) and both of $\Delta f_{1.4\,eV}>0$ and $\Delta\gamma_{1.4\,eV}>0$ (green curve) to the shape of $\Delta R/R$ .

(iv) The change in reflectivity below 1 eV from negative at $t_d$=0 ps (A) to positive at $t_d$=0.2 ps (F) is mainly due to a decrease in $\gamma_{MIR}$, i.e., the change from $\Delta\varepsilon_2<0$ ($t_d$=0 ps; A, due to $\Delta\gamma_{MIR}>0$) to $\Delta\varepsilon_2>0$ ($t_d$=0.2 ps; F, due to $\Delta\gamma_{MIR}<0$) below 0.5 eV [orange dashed curve in Figs. S4(c), (f)].

It may be difficult to visualize the origin of $\Delta R/R<0$ below 1 eV at $t_d$=0 ps from Figs.S4(b),(c). Figure S5(b) shows the contribution to the $\Delta R/R$ spectrum at $t_d$=0 ps from an increase in $f_{1.4\,eV}$. It is clear that : v) the negative $\Delta R/R$ below 1 eV ($t_d$=0 ps) is dominated by the increase in $f_{1.4\ \mathrm{eV}}$ (while the sub-peak structure at 1.5 eV is attributed to the increased $\gamma_{1.4}$).

| | | Steady state | $t_d$ = 0 ps | $t_d$ = 0.2 ps |
|---|---|---|---|---|
| Drude response | $\gamma_D$ [eV] (SC/N) | 0 | 0/0.010 | 0/0.004 |
| | $f_D$(×$10^{-2}$) | 2.80 | 2.369/0.742 | 2.119/1.001 |
| Mid-IR band | $\gamma_{MIR}$ | 0.88 [eV] | +0.52 % | -0.91 % |
| | $f_{MIR}$ | 6.96 (×$10^{-2}$) | +0.42 % | +0.10 % |
| 1.4 eV band | $\omega_{1.4\ eV}$ | 1.4 [eV] | 0 % | -0.93 % |
| | $\gamma_{1.4\ eV}$ | 0.50 [eV] | +13 % | 13 % |
| | $f_{1.4\ eV}$ | 3.2(×$10^{-3}$) | +14 % | +19 % |
| ZRS band | $f_{ZRS}$ | 1.3(×$10^{-3}$) | -11 % | -10 % |
| CT band | $f_{CT}$ | 9.51(×$10^{-3}$) | -0.96 % | -0.95 % |

**Table S2 Parameters used for DLM analysis of Δ*R/R* spectra (10 K)**. "0/0.010" ($t_d$=0 ps) and "0/0.004" ($t_d$=0.2 ps)" for the Drude response are $\gamma_D$ of the SC and N components, respectively. Here, we allowed variations of oscillator strengths ($f$) and scattering rates ($\gamma$) only, for the Drude response ($f_D$, $\gamma_D$), Mid-IR band ($f_{MIR}$, $\gamma_{MIR}$), 1.4 eV band ($f_{1.4\ eV}$, $\gamma_{1.4\ eV}$), ZRS band ($f_{ZRS}$, $\gamma_{ZRS}$) and CT band ($f_{CT}$, $\gamma_{CT}$). Just in the case of the 1.4 eV band we also allowed a variation of the central frequency ($\Delta\omega_{1.4\ eV}$).

**295 K**

Figures S6(a)-(c) show the spectrum in the normal state at 295 K and the results from the DLM analysis, presented in the same format as Figs. S4(a)-(c). The parameters used for the fitting are shown in Table S3. The contribution of each oscillator considered in the DLM to the structure of the Δ*R/R* spectrum in the normal state can be summarized as follows:

(i) Below 1 eV, the contribution to the decrease in reflectivity becomes larger at lower energies (A). This structure arises from the increased $\gamma_D$ of the Drude component (brown dashed curve).

(ii) The decrease in reflectivity peaking at 0.5 eV (B) and the increase in reflectivity up to 2.5 eV (C) are mainly due to the increased $\gamma_{MIR}$ and $f_{MIR}$ (orange dashed curve).

(iii) The slight decrease in reflectivity observed between 2.5 and 4 eV (D) is due to the

shift of the peak energy of the CT band, $\Delta\omega_{CT}$, (Hartree shift) (violet dashed curve).

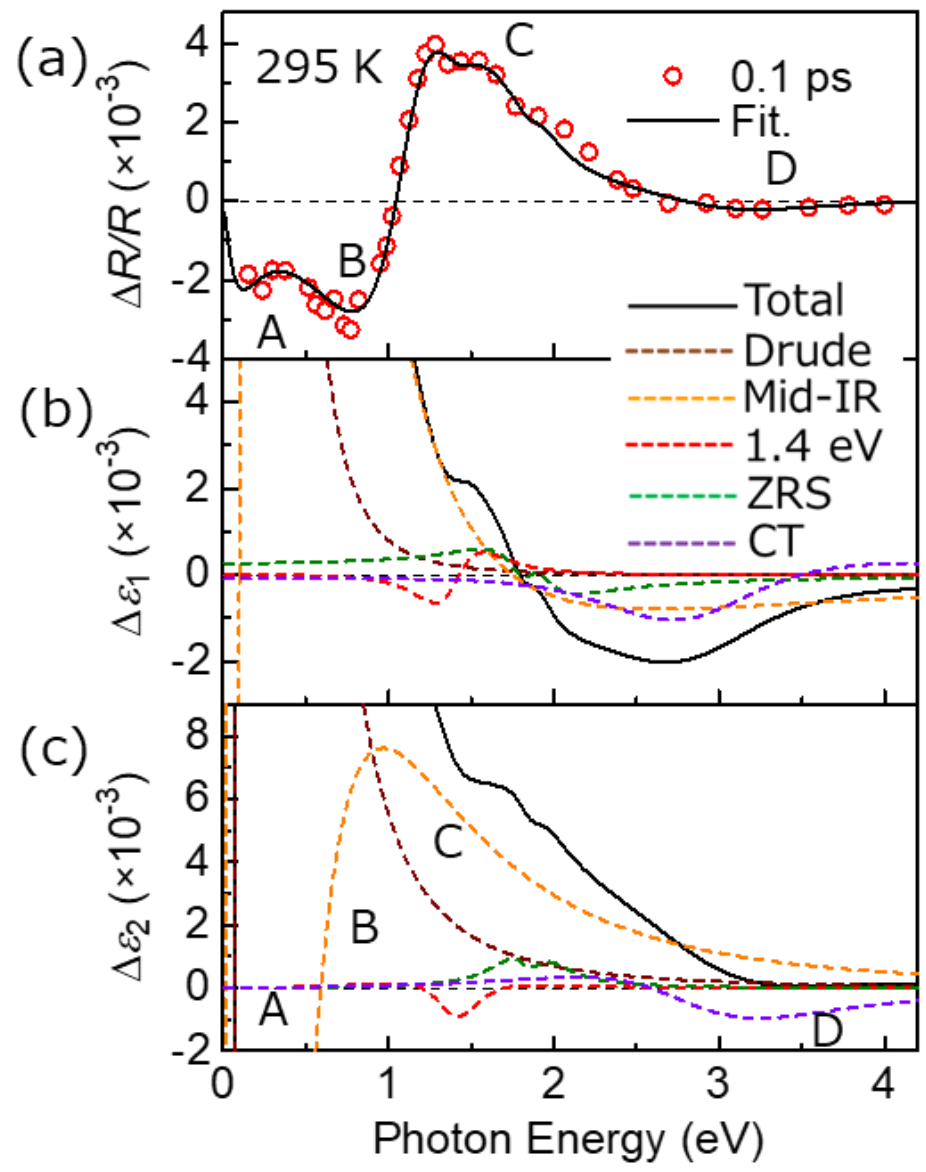

**Fig. S6 Details of spectral analysis (295 K).** (a) Results of the analysis for the $\Delta R/R$ spectrum measured at 295 K ($t_d$ = 0.1 ps). For A-D, refer to the text above. (b) Contribution of each oscillator to the shape of $\Delta\varepsilon_1$ (b) and that of $\Delta\varepsilon_2$ (c), i.e., the respective spectra were calculated by varying $f$ and $\gamma$ of a single focused oscillator. See text.

| | | Steady state | $t_d$ = 0.1 ps |
|---|---|---|---|
| Drude response | $\gamma_D$ | 0.07 [eV] | +5.6 % |
| Mid-IR band | $\gamma_{MIR}$ | 0.89 [eV] | +0.71 % |
| | $f_{MIR}$ | 7.33 ($\times 10^{-2}$) | +0.30 % |
| 1.4 eV band | $\gamma_{1.4\,eV}$ | 0.50 [eV] | +0.40 % |
| ZRS band | $\gamma_{ZRS}$ | 0.30 ($\times 10^{-2}$) | +0.67 % |
| | $f_{ZRS}$ | 0.15 ($\times 10^{-2}$) | +0.83 % |
| CT band | $\omega_{CT}$ | 2.93 [eV] | -0.041% |
| | $f_{CT}$ | 10.01 ($\times 10^{-2}$) | -0.025 % |

**Table S3 Parameters used for DLM analysis of $\Delta R/R$ spectrum (295 K)**. We allowed variations of the scattering rates ($\gamma$) only, for the Drude response ($\gamma_D$) and 1.4 eV band ($\gamma_{1.4\ eV}$), and those of the oscillator strengths ($f$) and scattering rates ($\gamma$) for the mid-IR band ($f_{MIR}$, $\gamma_{MIR}$) and ZRS band ($f_{ZRS}$, $\gamma_{ZRS}$). In the case of the CT band we allowed variations of the oscillator strength ($f_{CT}$) and central frequency ($\omega_{CT}$) .

## Supplemental material (6)

### Analysis of $\Delta R/R$ spectrum using multilayer model

The difference in the penetration depths [$L_p$=1/α (α is the absorption coefficient)] between the pump and probe lights may affect the shape of the $\Delta R/R$ spectra [72]. Since our measurements using 100 fs pulses cover a broad probe energy range (0.1–4.1 eV), we have to confirm that the effect of the mismatch in $L_p$ is critical or not. Figure S7(a) shows the $L_p$ spectrum in the optimally doped YBCO at 10 K. The value of $L_p$ varies between 70 and 160 nm at the probe energies between 0.1 and 4 eV, while that of $L_p$ of the pump light (of 0.9 eV) is ~90 nm.

To investigate the effects of this mismatch [as schematically shown in Fig. S7(b)], the $\Delta R/R$ spectra of the photoexcited state have been calculated using the multilayer model. The refractive index depending on the sample depth is assumed to be

$$\tilde{n}(\omega, z) = \tilde{n}_{st}(\omega)\{1 - \exp(-\alpha z)\} + \tilde{n}_{ex}(\omega)\exp(-\alpha z) \qquad \text{(eq. s2)}$$

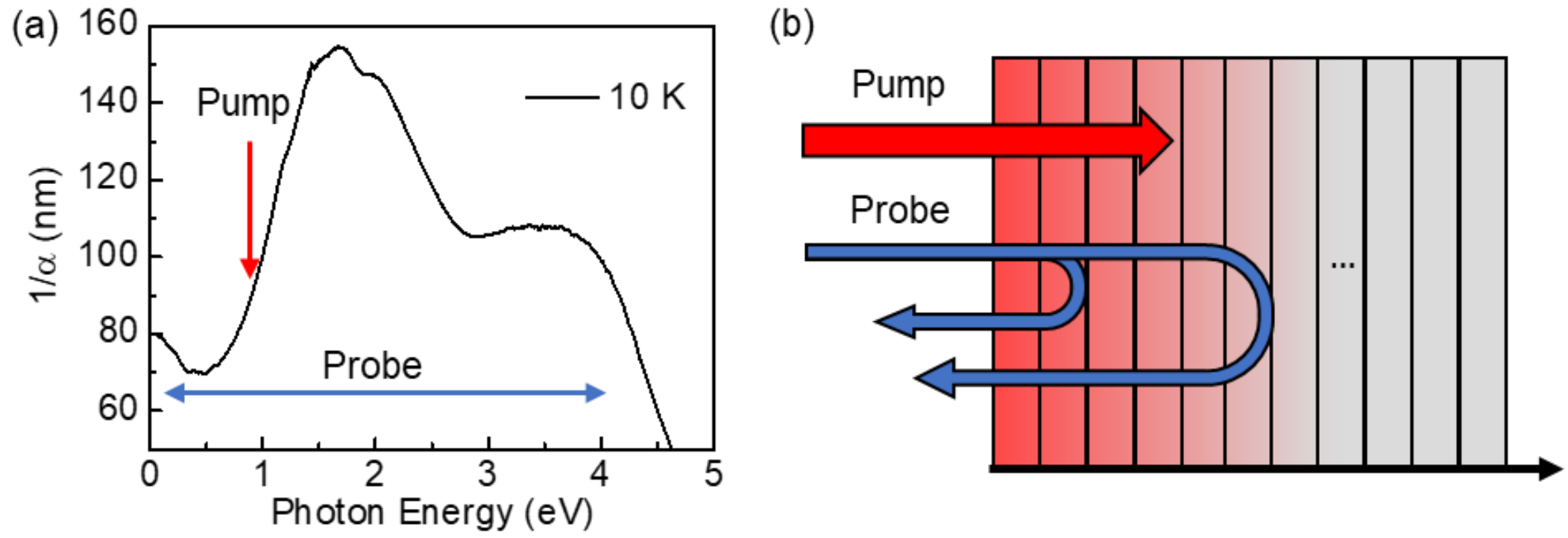

**Fig. S7 Difference in penetration depths between pump and probe lights.** (a) 1/α ($L_p$) spectrum in the optimally doped YBCO at 10 K. (b) Schematic illustration of the difference in the penetration depths ($L_p$=1/α ) between the pump and probe lights

, where the complex refractive indices at the sample surface in the ground and photoexcited states are $\tilde{n}_{st}(\omega)$ and $\tilde{n}_{ex}(\omega)$, respectively, and the z-axis is defined to be in the direction of the sample depth. The optical constants are assumed to be uniform within each layer, and the complex refractive index of the $i$-th layer ($z = z_i$) is calculated using (eq. s2). The thickness of each layer is $t$=0.5 nm, and the total number of layers is $N$=1000.

Figure S8 shows the $\Delta R/R$ spectrum at $t_d$=0.2 ps in the SC state and the results of the DLM analysis, with (red curve: "Multilayer") and without (blue curve, "Bulk") using the multilayer model. Thus, the DLM analysis can well reproduce the experimental results regardless of whether the multilayer model is used. The fitting parameters are summarized in Table S4. Using the multilayer model in the analysis affects the absolute values of the parameters of the DLM. Specifically, depending on the degree of mismatch in the absorption coefficients, using the multilayer model can increase the values of parameters by up to a factor of two. However, no significant changes were seen in the time evolution of the parameters.

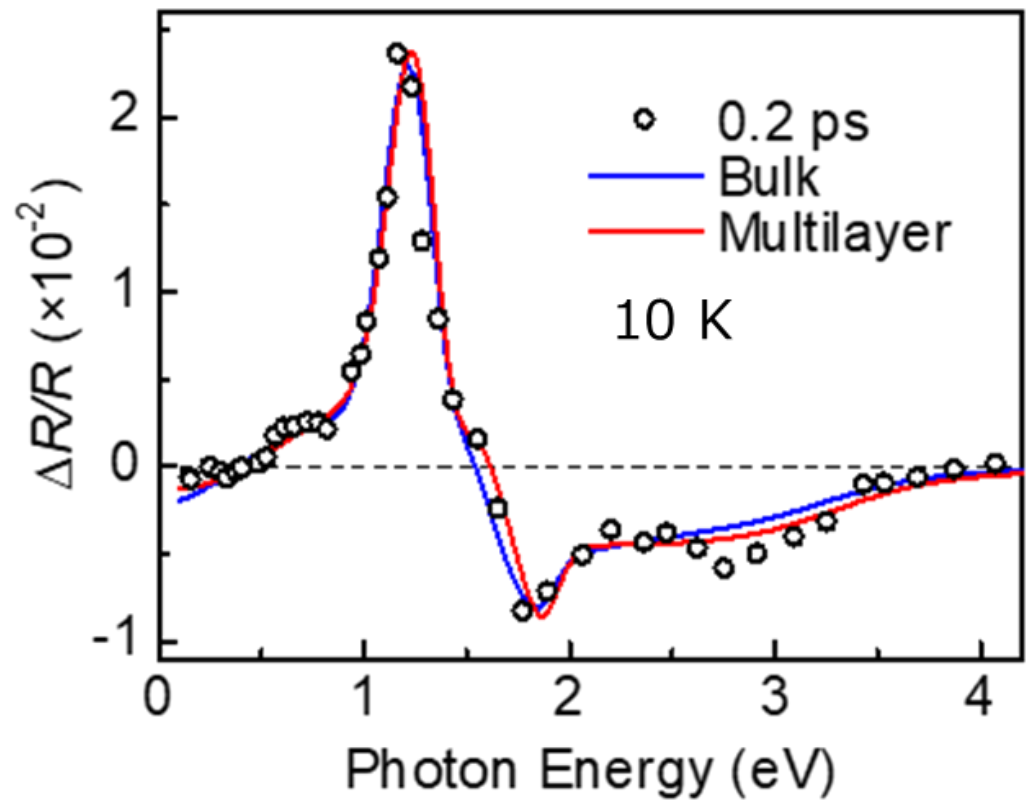

**Fig. S8 Analysis using multi-layer model.** $\Delta R/R$ spectrum at $t_d$=0.2 ps in the SC state (10 K) and the results of the DLM analysis, with (red curve: "Multilayer") and without (blue curve, "Bulk") using the multilayer model.

| | | Steady state | Multilayer | Bulk |
|---|---|---|---|---|
| Drude response | $\gamma_D$ [eV] | 0 | 0/0.004 | 0/0.004 |
| | $f_D$ ($\times 10^{-2}$) | 2.80 | 2.119/1.001 | 2.119/0.998 |
| Mid-IR band | $\gamma_{MIR}$ | 0.88 [eV] | -0.91 % | -0.57 % |
| | $f_{MIR}$ | 6.96 ($\times 10^{-2}$) | +0.10 % | 0 % |
| 1.4 eV band | $\omega_{1.4\ eV}$ | 1.4 [eV] | -0.93 % | -0.64 % |
| | $\gamma_{1.4\ eV}$ | 0.50 [eV] | +13 % | +8.4 % |
| | $f_{1.4\ eV}$ | 3.2($\times 10^{-3}$) | +19 % | +16 % |
| ZRS band | $f_{ZRS}$ | 1.3 ($\times 10^{-3}$) | -10 % | -5.8 % |
| CT band | $f_{CT}$ | 9.51($\times 10^{-2}$) | -0.95 % | -0.74 % |

**Table S4 Parameters for DLM analysis of $\Delta R/R$ spectrum with (Multilayer) and without (Bulk) using multilayer model ($t_d$=0.2ps at 10 K).** "0/0.004" for the Drude response are $\gamma_D$ of the SC and N components, respectively.

**Supplemental material (7)**

**Photoinduced change in loss function induced by increase in $f$ of 1.4 eV band**

Figure S9 shows the imaginary part of the loss function (Im[$1/\varepsilon$]) spectrum obtained from the analysis of the steady-state spectrum (black line), and that with increased $f_{1.4\ \mathrm{eV}}$ (red line). The plasma peak of the loss function becomes smaller with the increase in $f_{1.4\ \mathrm{eV}}$. This change caused by the increase in $f_{1.4\ \mathrm{eV}}$ is consistent with the change in the loss function from 20 K to 300 K reflecting the recovery of the Umklapp-scattering effect [29, 30].

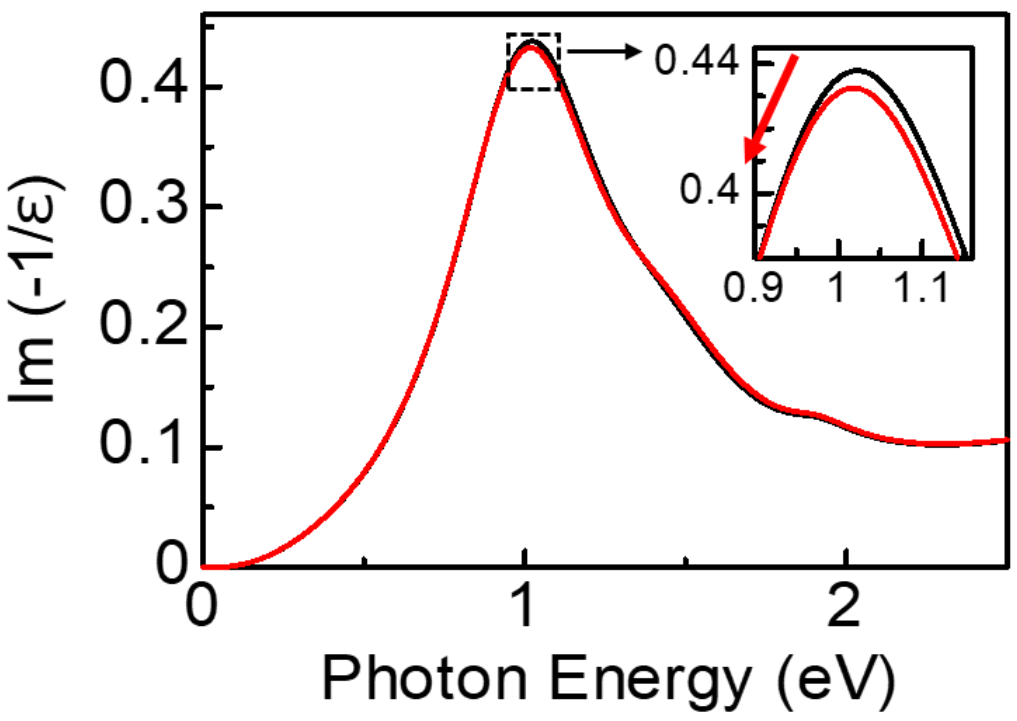

**Fig. S9 Dieelectric loss function spectra.** Imaginary part of the loss function (Im[$1/\varepsilon$]) spectrum obtained from the analysis of the steady-state spectrum (black line), and that with increased $f_{1.4\ \mathrm{eV}}$ (red line) calculated based on the DLM. The inset shows an enlarged view of the region near the 0.9–1.1 eV peak.

**Supplemental material (8)**

**Time evolutions of $\Delta\gamma$, $\Delta f$ and $\Delta\omega$ in normal state**

Figure S10(a) shows the $\Delta R/R$ spectrum at $t_{\mathrm{d}}$ = 0.1 ps (red circles) and the fitting curve (solid red line) analyzed by the DLM (295 K). The time evolutions of the fitting parameters ($\Delta\gamma$, $\Delta f$) are shown in Fig. S10(b–i): Mid-IR band (orange: b, c); 1.4 eV band (red: d, e); ZRS band (green: f, g); and CT band (violet: h, i). The time evolution of the peak energy of the CT band ($\Delta\omega_{\mathrm{CT}}$) is shown as the violet filled circles in Fig. S10(i).

Contrary to the results in the SC phase, $\gamma_{\mathrm{MIR}}$ increases instantaneously with optical excitation. Furthermore, $f_{1.4\ \mathrm{eV}}$ remains constant, since the Umklapp-scattering effect is not suppressed in the normal state in contrast to the SC state. Additionally, $f_{\mathrm{ZRS}}$ increases, unlike in the SC state. This corresponds to the Hartree shift occurring in opposite directions in the SC and normal states. All of these responses decay within a few hundred femtoseconds, exhibiting the typical optical response of a metal due to the rise and relaxation in electron temperature.

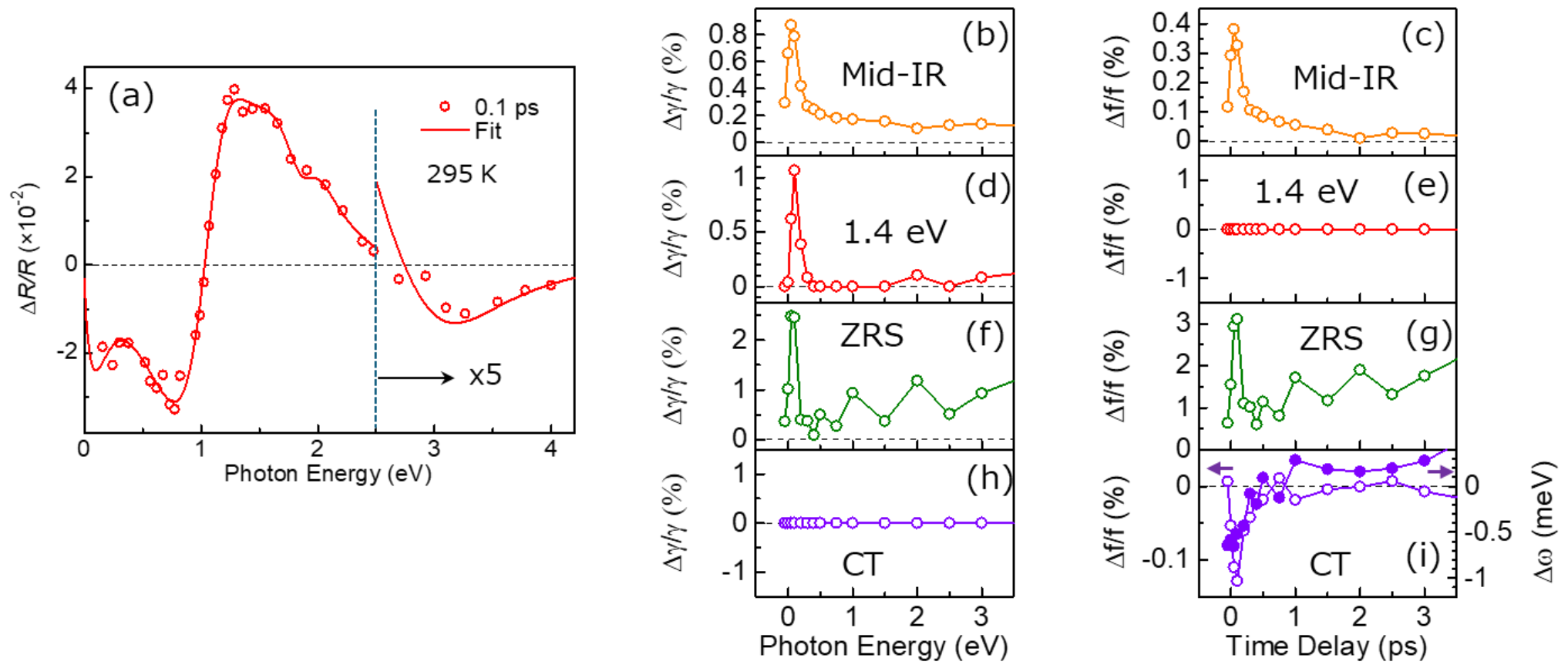

**Fig. S10 Results of DLM analysis (295 K).** (a) The $\Delta R/R$ spectrum at $t_d = 0.1$ ps (red circles, 295 K) and the fitting curve (solid red line) analyzed by the DLM. (b)-(i) The time evolutions of the fitting parameters ($\Delta\gamma$, $\Delta f$) are shown: $\Delta\gamma_{\mathrm{MIR}}$, $\Delta f_{\mathrm{MIR}}$ (b, c); $\Delta\gamma_{1.4\ \mathrm{eV}}$, $\Delta f_{1.4\ \mathrm{eV}}$ (d, e); $\Delta\gamma_{\mathrm{ZRS}}$, $\Delta f_{\mathrm{ZRS}}$ (f, g); and $\Delta\gamma_{\mathrm{CT}}$, $\Delta f_{\mathrm{CT}}$ (h, i). The time evolution of the peak shift (right axis $\Delta\omega_{\mathrm{CT}}$) of the CT band is also shown as the violet closed circles in (i).

**Supplemental material (9)**

**Relation between hole density in *p* orbital and stability of SC**

It is known that the higher the transition temperature at optimal doping, the greater the hole density in the *p* orbital $n_p$, according to the measurements of nuclear quadrupole (NQR) and nuclear magnetic resonance (NMR) [66-68]. Furthermore, it is known that $n_p$ increases below the transition temperature [69] on the basis of the results of polarized X-ray absorption measurements.

**Supplemental material (10)**

**Possible photoinduced Hartree shift estimated from photon density**

The magnitude of the possible Hartree shift is estimated to be 2 meV if holes are efficiently transferred by photoexcitation, i.e., if each absorbed photon (9.8 x $10^{-4}$ /site) is used to transfer a hole between the different orbitals. Because of the low excitation efficiency of the CT band by the 0.89 eV pumping, the actual energy shift should be smaller than ca. 1 meV. This Hartree shift predicted from the excitation density and absorption coefficient is smaller than the energy resolution (ca. 30 meV) of our $\Delta R/R$ measurements using 100 fs pulses. However, since $\Delta R/R$ measurements directly detect photoinduced changes (a modulation technique), the energy resolution of this measurement is determined by the sensitivity of the differential reflectivity. At room temperature, the decrease in reflectivity observed around 3 eV can be explained by a Hartree shift of ca. 0.7 meV [see Fig. S10(i)]. The instantaneous response of the red shift ($\Delta\omega_{CT}<0$) of the CT band is shown by violet filled circles in Fig. S10(i). In the superconducting state, however, a prominent bleaching of the CT band overlap occurs in this energy region, which makes the presence of the Hartree shift less clear.